\documentclass[article]{emulateapj}
\usepackage{natbib}
\newcommand{\hi}{\mbox{\rm \ion{H}{1}}} 

\newcommand{\htwo}{\mbox{\rm H$_2$}}

\newcommand{\xcounits}{\mbox{cm$^{-2}$ (K km s$^{-1}$)$^{-1}$}}
\newcommand{\alphaunits}{\mbox{M$_\odot$~pc$^{-2}$~(K km s$^{-1}$)$^{-1}$}}

\newcommand{\xco}{\mbox{$X_{\rm CO}$}}
\newcommand{\alphaco}{\mbox{$\alpha_{\rm CO}$}}

\shorttitle{\alphaco\ From Dust In The Local Group} \shortauthors{Leroy et al.}
\slugcomment{Draft Version}

\begin{document}
\title{The CO-to-H$_2$ Conversion Factor From Infrared Dust Emission
  Across the Local Group}

\author{Adam K. Leroy\altaffilmark{1,12}, Alberto
  Bolatto\altaffilmark{2}, Karl Gordon\altaffilmark{3}, Karin
  Sandstrom\altaffilmark{4}, Pierre Gratier\altaffilmark{5}, Erik
  Rosolowsky\altaffilmark{6}, Charles W. Engelbracht\altaffilmark{7},
  Norikazu Mizuno\altaffilmark{8}, Edvige Corbelli\altaffilmark{9},
  Yasuo Fukui\altaffilmark{10}, Akiko Kawamura\altaffilmark{10}}

\altaffiltext{1}{National Radio Astronomy Observatory, 520 Edgemont
  Road, Charlottesville, VA 22903, USA}

\altaffiltext{2}{Department of Astronomy, University of Maryland, College
  Park, MD 20742} 

\altaffiltext{3}{Space Telescope Science Institute, 3700 San Martin Drive, Baltimore, MD 21218, USA} 

\altaffiltext{4}{Max-Planck-Institut f{\" u}r Astronomie, D-69117 Heidelberg,
  Germany} 

\altaffiltext{5}{Laboratoire d'Astrophysique de Bordeaux, Universit\'{e}
  de Bordeaux, OASU, CNRS/INSU, 33271 Floirac, France}

\altaffiltext{6}{University of British Columbia Okanagan, 3333 University Way,
Kelowna, BC V1V 1V7, Canada}

\altaffiltext{7}{Steward Observatory, University of Arizona, Tucson,
  AZ 85721}

\altaffiltext{8}{National Astronomical Observatory of Japan, 2-21-1
  Osawa, Mitaka 181-8588 Tokyo, Japan}

\altaffiltext{9}{INAF-Osservatorio Astrofisico di Arcetri, Largo
  E. Fermi 5, 50125 Firenze, Italy}

\altaffiltext{10}{Department of Astrophysics, Nagoya University,
  Furocho, Chikusaku, Nagoya 464-8602, Japan}

\altaffiltext{11}{Hubble Fellow}

\begin{abstract}
We estimate the conversion factor relating CO emission to \htwo\ mass,
$\alpha_{\rm CO}$, in five Local Group galaxies that span
approximately an order of magnitude in metallicity --- M~31, M~33, the
Large Magellanic Cloud (LMC), NGC~6822, and the Small Magellanic Cloud
(SMC). We model the dust mass along the line of sight from infrared
(IR) emission and then solve for the $\alpha_{\rm CO}$ that best
allows a single gas-to-dust ratio ($\delta_{\rm GDR}$) to describe
each system. This approach remains sensitive to CO-dark envelopes
\htwo\ surrounding molecular clouds. In M~31, M~33, and the LMC we
find $\alpha_{\rm CO} \approx 3$--$9$~\alphaunits , consistent with
the Milky Way value within the uncertainties. The two lowest
metallicity galaxies in our sample, NGC~6822 and the SMC
($12+\log({\rm O/H})\approx 8.2$ and $8.0$), exhibit a much higher
$\alpha_{\rm CO}$. Our best estimates are $\alpha_{\rm CO}^{\rm
  NGC6822}\approx 30$~\alphaunits\ and $\alpha_{\rm CO}^{\rm SMC}
\approx 70$~\alphaunits . These results are consistent with the
conversion factor becoming a strong function of metallicity around
$12+\log ({\rm O/H}) \sim 8.4-8.2$. We favor an interpretation where
decreased dust-shielding leads to the dominance of CO-free envelopes
around molecular clouds below this metallicity.
\end{abstract}


\section{Introduction}
\label{sec:intro}

In the local universe, stars form out of clouds made of molecular
hydrogen (\htwo ) \citep[e.g.,][]{FUKUI10}. Understanding the
processes that lead to star formation on large scales requires
measuring the mass and distribution of this gas. Unfortunately, H$_2$
lacks a dipole moment, most molecular gas is found under conditions
too cold to excite quadrupole emission, and the high opacity of
molecular clouds prevents UV absorption studies from probing the bulk
of the gas. Estimates of a galaxy's \htwo\ distribution therefore rely
on indirect tracers, most commonly the lower rotational transitions of
CO. The conversion between CO intensity and \htwo\ abundance has been
the topic of a great deal of investigation. In particular, the effect
of metallicity and the local radiation field on the CO-to-H$_2$ mass
conversion factor, \alphaco \footnote{We work with $\alpha_{\rm CO}$,
  the conversion from integrated CO intensity to mass of molecular
  gas. A linear scaling relates $\alpha_{\rm CO}$ to \xco , the
  conversion from integrated CO intensity to column density of \htwo
  . Including helium, $\xco$~\xcounits $= 4.6 \times 10^{19}
  \alpha_{\rm CO}$~\alphaunits .}, have been studied for more than two
decades.

Infrared (IR) dust emission is a powerful tool to address this
problem. Dust is observed to be well-mixed with gas
\citep[e.g.,][]{BOHLIN78, BOULANGER96} and may be mapped by its
emission at IR wavelengths \citep[e.g.,][]{SCHLEGEL98}. By comparing
IR emission, CO, and \hi\ one can constrain
\alphaco\ \citep[e.g.,][]{THRONSON88B,ISRAEL97}. The procedure is to
estimate the dust mass from IR emission, measure atomic gas (\hi ) and
CO emission over a matched area, and then assume that the total gas
mass ($\htwo + \hi$) is proportional to the dust mass, with the two
related by a fixed gas-to-dust ratio. With measurements that span a
range of relative CO, \hi , and dust masses, it is possible to
simultaneously constrain the gas-to-dust ratio and \alphaco . Key
advantages of this approach are that it remains sensitive to any
CO-free envelopes of \htwo\ and that the calibration of $\alpha_{\rm
  CO}$ is pinned to the \hi\ within the system being studied.

This exercise has been applied to several dwarf irregular galaxies
\citep[][Bolatto et al. 2010, in
  preparation]{ISRAEL97,ISRAEL97B,LEROY07,LEROY09,GRATIER10}. In the
low-metallicity Small Magellanic Cloud, the results suggest a very
large $\alpha_{\rm CO} \sim 90$--$270$~\alphaunits . An analogous
application to the Milky Way yields $\alpha_{\rm CO} \sim
4$--$9$~\alphaunits\ \citep{BLOEMEN90, DAME01}, roughly compatible
with determinations of \alphaco\ from fitting the diffuse $\gamma$-ray
background \citep[][]{STRONG96,ABDOXCO}.

Most studies of $\alpha_{\rm CO}$ have focused on a single galaxy or
high latitudes in the Milky Way, where confusion is
minimal. \citet{ISRAEL97} worked with a varied galaxy sample but since
his work the available IR and CO data for nearby galaxies have
improved dramatically. This is largely thanks to the {\em Spitzer}
Space Telescope, which recently finished its cool mission and produced
high quality maps of a number of Local Group galaxies. This is
therefore a natural time to apply this technique to measure
\alphaco\ across the Local Group in a self-consistent way.

In this paper, we combine maps of CO, \hi , and IR emission to
estimate \alphaco\ in five Local Group galaxies: the massive spiral
M~31, the dwarf spiral M~33, the dwarf irregular NGC~6822, and the
Large and Small Magellanic Clouds (the LMC and SMC). By treating all
five systems self-consistently, we minimize uncertainty in the "zero
point" of the approach. With the {\em Herschel} mission now underway
and the execution of several complementary CO and \hi\ surveys, this
approach should be readily extensible to many nearby galaxies in the
next few years.

\section{The Model}
\label{sec:model}

We and assume that dust and gas are linearly related by a gas-to-dust
ratio, $\delta_{\rm GDR}$, so that

\begin{equation}
\delta_{\rm GDR}~\Sigma_{\rm dust} = \Sigma_{\rm H2} + \Sigma_{\rm HI}~.
\end{equation}

\noindent where $\Sigma_{\rm dust}$, $\Sigma_{\rm H2}$, and
$\Sigma_{\rm HI}$ are the mass surface densities of dust, \htwo , and
\hi\ along a line of sight. Substituting $\Sigma_{\rm H2} =
\alpha_{\rm CO}~I_{\rm CO}$ we have

\begin{equation}
\label{eq:model}
\delta_{\rm GDR}~\Sigma_{\rm dust} = \alpha_{\rm CO}~I_{\rm CO} + \Sigma_{\rm
  HI}~,
\end{equation}

\noindent where the CO-to-H$_2$ conversion factor, $\alpha_{\rm CO}$,
and gas-to-dust ratio, $\delta_{\rm GDR}$, are unknown and
$\Sigma_{\rm dust}$, $I_{\rm CO}$, and $\Sigma_{\rm HI}$ are
measured. After assembling $\Sigma_{\rm dust}$, $I_{\rm CO}$, and
$\Sigma_{\rm HI}$ over many lines of sight in a region, we will use
these data to solve for $\alpha_{\rm CO}$ that best allows a single
$\delta_{\rm GDR}$ to describe the data.

\subsection{$\alpha_{\rm CO}$: Definitions and Scales}

The literature contains several working definitions of
  $\alpha_{\rm CO}$ that apply to different scales or phases of the
  gas. In this paper and Equation \ref{eq:model} we define
  $\alpha_{\rm CO}$ as the factor to convert from CO emission to {\em
    total} molecular gas mass on scales larger than individual
  clouds. Under this definition, $\alpha_{\rm CO}$ includes any H$_2$
  associated with C$^{+}$ in the outer, poorly shielded parts of
  clouds as well as gas immediately mixed with CO. Indeed, our goal is
  to measure whether such envelopes become dominant below some
  metallicity. Because this definition integrates over cloud
  structure, it is possible to derive a single $\alpha_{\rm CO}$ for a
  whole galaxy or part of a galaxy, and to view that $\alpha_{\rm CO}$
  as a function of large-scale environmental factors. We choose this
  definition of $\alpha_{\rm CO}$ because it is directly applicable to
  CO measurements of distant galaxies on kiloparsec scales.

This is distinct from the ratio of CO to H$_2$ only for molecular
  gas mixed with CO. Dynamical measurements using CO emission at high
  spatial resolution may mainly probe this quantity with limited
  sensitivity to H$_2$ in an extended envelope not mixed with CO. The
  difference between this quantity and the quantity that we study has
  caused some confusion, leading to apparent contradictions between
  IR-based measurements and dynamical measurements. In fact these may
  be largely attributed to the different regions being probed (Section
  \ref{sec:discussion}).

Similarly, we do {\em not} define $\alpha_{\rm CO}$ as the ratio
  of H$_2$ to CO along a pencil beam. We have no reason to expect that
  this quantity, or any ratio that places many elements across a
  cloud, remains reasonably constant across part of a galaxy. To the
  contrary comparisons of dust and CO emission imply dramatic
  variations in the CO-to-H$_2$ ratio within individual clouds
  \citep{PINEDA08}.

Our spatial resolution ranges from 45--180~pc (Table
  \ref{tab:data}). For these resolutions, giant molecular clouds will
  mostly lie within one or two resolution elements
  \citep[e.g.,][]{HEYER09}. This is ideal for our definition of
  $\alpha_{\rm CO}$. We wish to integrate over the structure of these
  clouds to make a single $\alpha_{\rm CO}$ a more appropriate
  assumption.

\section{Data}
\label{sec:data}

\begin{deluxetable}{l c c c}
  \tabletypesize{\small} 
  \tablewidth{0pt} 
  \tablecolumns{4}
  \tablecaption{\label{tab:data} Targets}
  \tablehead{\colhead{System} &
    \colhead{Metallicity\tablenotemark{a}} &
    \multicolumn{2}{c}{Resolution} \\
    & 12+$\log ({\rm O/H})$ & ($\arcsec$) & (pc) }
  \startdata
  M~31 & & \\
  ... inner & 9.0 & $45$ & $170$ \\
  ... north & 8.7 & $45$ & $170$ \\
  ... south & 8.7 & $45$ & $170$ \\
  M~33 & & \\
  ...inner & 8.33 & $45$ & $180$ \\
  ...outer & 8.27 & $45$ & $180$ \\
  LMC & 8.43 & $240$ & $60$ \\
  NGC 6822 & 8.2 & $45$ & $110$ \\
  SMC & & \\
  ... west & 8.02 & $156$ & $45$ \\
  ... east & 8.02 & $156$ & $45$ \\
  ... north & 8.02 & $156$ & $45$ \\
  \enddata
  \tablenotetext{a}{{\sc References:} {\sc M~31} --- \citet{YIN09}
  (compilation); {\sc M~33} --- \citet{ROSOLOWSKY08B}; {\sc LMC} and
  {\sc SMC}--- \citet{DUFOUR84,KELLER06}; {\sc NGC 6822} ---
  \citet{ISRAEL97B} (compilation). Adopted to match \citet{BOLATTO08}
  where possible.}
\end{deluxetable}

To carry out this experiment we require maps of IR (to estimate
$\Sigma_{\rm dust}$), CO, and \hi\ emission. Such maps have been
published for M~31, M~33, NGC~6822, the LMC and the SMC. We refer to the
original papers for details of the observations, reduction, and data.

For M~31, we use the CO map taken by \citet{NIETEN06} using the IRAM
30m telescope. This map has already been masked by \citet{NIETEN06}
and so contains only positive signal. We trace \hi\ using the 21cm map
of \citet{BRINKS84}, obtained with the Westerbork Synthesis Radio
Telescope (WSRT). \citet{GORDON06} present {\em Spitzer} maps at 24,
70, and 160$\mu$m.

For M~33, we take CO data from \citet{ROSOLOWSKY07B}, which combines
BIMA \citep{ENGARGIOLA03} and FCRAO data \citep{HEYER04}. We use the
WSRT map by \citet{DEUL87}. We use {\em Spitzer} data reduced
following \citet{GORDON05} and presented by \citet{VERLEY07} and
\citet{TABATABAEI07}.

We use the first CO map of the LMC obtained by NANTEN \citep{FUKUI99},
the Australia Telescope Compact Array (ATCA) + Parkes \hi\ map of
\citet{KIM98,KIM03}, and IR maps from the {\em Spitzer} SAGE legacy
program \citep{MEIXNER06,BERNARD08}.

We also use the NANTEN CO map of the SMC \citep{MIZUNO01}, the ATCA +
Parkes \hi\ map by \citet{STANIMIROVIC99}. The SMC IR maps are a
combination of data from the SAGE-SMC legacy program \citet[][and in
  prep.]{GORDON09} and the S3MC survey \citep{BOLATTO07}.

For NGC~6822, we use the IRAM~30m CO map by \citet{GRATIER10}, the
SINGS {\em Spitzer} maps presented by \citet{CANNON06}, and the VLA
\hi\ map of \citet{DEBLOK03}. \citet{GRATIER10} mapped the CO
$J=2\rightarrow1$ line for this galaxy, while the rest of our maps are
of CO $J=1\rightarrow0$. We follow \citet{GRATIER10} in assuming a
line ratio of $0.7$; we phrase all of our results in terms of CO
$J=1\rightarrow0$ intensity assuming this ratio. We mask the CO map,
keeping only emission above $\sim 3\sigma$ at the original 15$\arcsec$
resolution.

For each galaxy we convolve all data to the resolution of the coarsest
data set and align them on a common astrometric grid. In M~31, M~33,
and NGC~6822 we are limited by the resolution of the {\em Spitzer} 160
$\mu$m data and so convolve all data to have a $45\arcsec$ (FWHM)
Gaussian PSF. In the LMC the NANTEN CO data limit our resolution to $>
2\arcmin .6$; because signal-to-noise in the CO map is also a concern,
we convolve all data to $4\arcmin$ resolution. The NANTEN CO data also
set the resolution in the SMC, where we convolve all data to $2\arcmin
.6$ resolution.

In the LMC and SMC we subtract a foreground from the IR
maps. Following \citet{BOT04}, we remove a scaled version of the Milky
Way \hi\ over these lines of sight \citep[for the exact approach
  see][]{LEROY09}. In M~33 and M~31 the reduction imposes the
condition that the intensity is 0 away from the galaxy, making a
cirrus subtraction unnecessary. In NGC~6822, \citet{CANNON06} already
removed a Galactic foreground.

The statistical uncertainty in the CO maps is about
$0.30$~K~km~s$^{-1}$ in M~31, $0.35$~K~km~s$^{-1}$ in M~33,
$0.30$~K~km~s$^{-1}$ in the LMC, $0.01$~K~km~s$^{-1}$ in NGC~6822, and
$0.08$~K~km~s$^{-1}$ in the SMC (though in each case this varies
somewhat with position). The statistical noise in the \hi\ maps is
very roughly $1$--$3$~M$_\odot$~pc$^{-2}$ with the uncertainty
dominated by imperfect knowledge of the \hi\ opacity and the
reconstruction of extended emission. The noise in the IR maps is $\sim
0.2$~MJy~sr$^{-1}$ at 70$\mu$m and $\sim 0.7$~MJy~sr$^{-1}$ at
160$\mu$m. The zero point in the IR maps is uncertain by $\sim
0.1$~MJy~sr$^{-1}$ at 70$\mu$m and $\sim 0.5$~MJy~sr$^{-1}$ at
160$\mu$m.

Throughout this paper IR intensity has units of MJy sr$^{-1}$,
color-corrected to the IRAS scale. \hi\ surface density has units of
M$_\odot$~pc$^{-2}$ and includes a factor of 1.36 to account for
helium. In the SMC and the LMC, \hi\ includes an opacity correction
based on \citet{DICKEY00}. In M~31, M~33, and NGC 6822 we have assumed
that the \hi\ is optically thin. CO $J=1\rightarrow0$ intensity has
units of K km s$^{-1}$ and is related to \htwo\ surface density, in
units of M$_\odot$~pc$^{-2}$, by a factor $\alpha_{\rm CO}$ that
includes helium (so that $\Sigma_{\rm H2} = \alpha_{\rm CO} I_{\rm
  CO}$); if $\xco = 2 \times 10^{20}$~\xcounits\ then $\alpha_{\rm CO}
= 4.4$~\alphaunits .

\section{Method}
\label{sec:method}

Our goal is to identify regions where both \hi\ and \htwo\ contribute
significantly to the interstellar medium (ISM), use the IR intensity
in these regions to estimate the dust column, and then harness the
assumption that gas and dust are linearly related to solve for
$\alpha_{\rm CO}$.

\subsection{Sampling and Target Region}
\begin{figure*}
\plotone{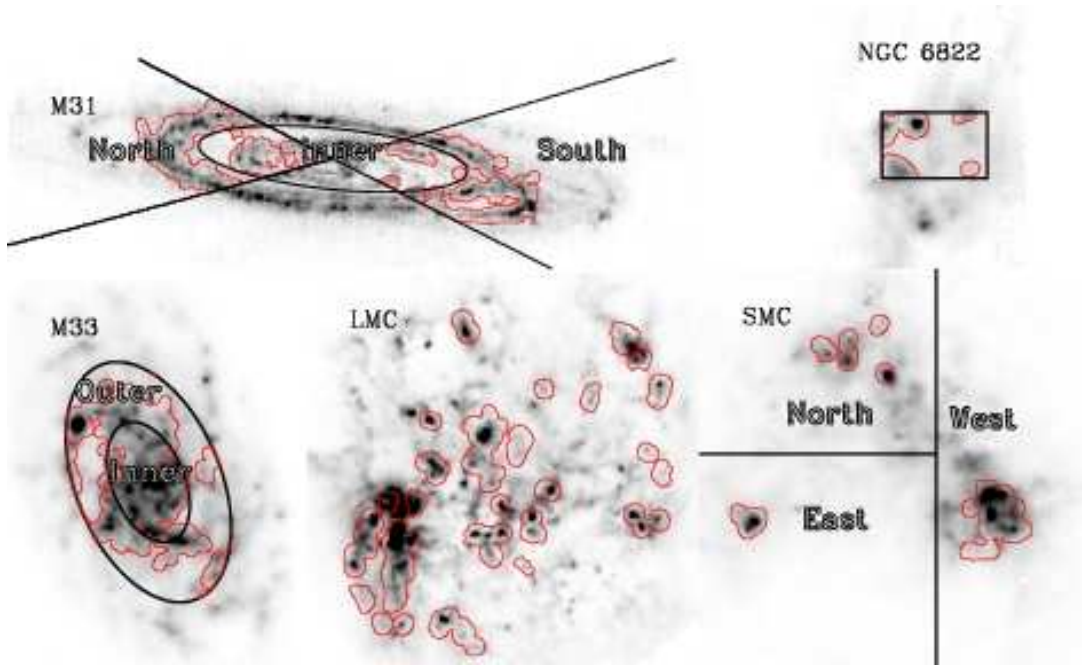}
\caption{\label{fig:sampling} Images of our targets at 160$\mu$m
  (grayscale) with the region considered enclosed by a red contour. In
  M~31, M~33, and the SMC black lines separate regions that we treat
  separately and labels indicate how we refer to these in the text.}
\end{figure*}

This experiment leverages our knowledge of \hi\ column to infer
$\alpha_{\rm CO}$. It thus works best in regions where both \hi\ and
\htwo\ are important to the ISM mass budget. If we target areas where
\hi\ dominates then $\alpha_{\rm CO}$ has little or no impact on
$\delta_{\rm GDR}$ while if we target areas with only \htwo\ then
$\alpha_{\rm CO}$ and $\delta_{\rm GDR}$ are degenerate.

For our targets and resolution, being dominated by \htwo\ is not a
concern for any reasonable $\alpha_{\rm CO}$. On the other hand each
target has large areas where the ISM is overwhelmingly \hi
. Especially in M~31 and M~33, much of the \hi\ is at large radius and
appears to have a different $\delta_{\rm GDR}$ from the inner galaxy
\citep[][]{NIETEN06}. We mostly avoid these \hi -only regions, instead
targeting the part of each galaxy near where CO is detected. This
gives us a range of total gas surface densities and relative
contributions by \htwo\ and \hi , allowing us to constrain
$\alpha_{\rm CO}$ and $\delta_{\rm GDR}$.

We define our target region by a $\approx 3\sigma$ intensity cut in
the convolved CO map, $I_{\rm CO} \geq 1$~K~km~s$^{-1}$ in M~31, M~33,
and the LMC, $I_{\rm CO} \geq 0.25$~K~km~s$^{-1}$ in the SMC, and
$I_{\rm CO} \geq 0.03$~K~km~s$^{-1}$ in NGC~6822. We reject small
regions (area $\lesssim$ a resolution element) as likely noise spikes
and then consider all area within about $1$~resolution element of the
remaining emission. Finally, we require a line of sight to have
significant IR emission to be included; this is $I_{160} \geq
5$~MJy~sr$^{-1}$ in NGC~6822, $I_{160} \geq 10$~MJy~sr$^{-1}$ in
all other targets. The result resembles loosely circling bright CO
emission by hand. Figure \ref{fig:sampling} shows the target regions
in contour on top of the 160$\mu$m map.

In M~31, M~33, and the SMC treating the whole galaxy at once causes
problems with our model, which assumes a constant $\alpha_{\rm CO}$
and $\delta_{\rm GDR}$ across the area studied. Previous work arrived
at similar conclusions. \citet{NIETEN06} observed $\delta_{\rm GDR}$
to be higher in the inner part of M~31 than the 10~kpc ring containing
most of the CO. In M~33, bright CO extends a fair distance out into
the disk, but based on radial profiles of \hi , CO, and $160\mu$m
intensity it is immediately clear that this galaxy, too, has a strong
radial gradient in $\delta_{\rm GDR}$. The SMC's CO emission is
clustered into three distinct regions that show evidence for local
variations in their $\delta_{\rm GDR}$, $\alpha_{\rm CO}$, and giant
molecular cloud (GMC) properties \citep{LEROY07,MUELLER10}.

To isolate regions with fixed $\alpha_{\rm CO}$ and $\delta_{\rm
  GDR}$, we separate each galaxy into several zones. We treat the
"inner" part of M~31 separately from the 10~kpc ring and further
divide the ring into a "north" and "south" part. We exclude
$60\arcdeg$ around the minor axis (in the plane of the galaxy) of M~31
(see \S \ref{sec:diffuse_hi}). We break M~33 into an "inner" zone
where $r_{\rm gal} < 2$~kpc and an "outer" zone where $2$~kpc~$<
r_{\rm gal} <$~4~kpc. We divide the SMC into three parts: a "west"
region, a "north" region, and an "east" region. Our regions in the SMC
deliberately exclude two clouds near the center of the galaxy
identified in \citet{LEROY07} to have high CO-to-IR ratios; including
these clouds leads to even worse solutions for this part of the
SMC. Black lines and labels in Figure \ref{fig:sampling} indicate the
divisions for each galaxy.

After the target regions are defined, we sample each map using a
hexagonal grid spaced by $1/2$ the resolution (i.e., $22.5\arcsec$ in
M~31, $20\arcsec$ in M~33, $2\arcmin$ in the LMC, $22.5\arcsec$ in
NGC~6822, and $1\arcmin .3$ in the SMC). This yields a matched,
approximately Nyquist-sampled set of measurements of $I_{70}$,
$I_{160}$, $I_{CO}$, and $\Sigma_{\rm HI}$.

\subsection{Estimating Dust Along the Line of Sight}

We follow the approach of \citet{DRAINE07A} and \citet{DRAINE07B} to
estimate the amount of dust along the line of sight. These papers
present models that can be used to estimate the dust mass and incident
radiation field from IR intensities. Following \citet{DRAINE07B} and
\citet{MUNOZMATEOS09}, we search a grid of models illuminated by
different radiation fields to find the best-fit dust mass surface
density, $\Sigma_{\rm dust}$, for each set of IR intensities. A slight
difference between our fits and those papers is that we quote the
geometric mean of the dust mass across the region of parameter space
where $\chi^2 < \chi^2_{\rm min} + 1$ (i.e., within 1 of the $\chi^2$
for the best-fit model). We present the results of our own direct
grid-search fits, but during analysis made extensive use of the the
work of \citet{MUNOZMATEOS09}, who parameterized the results of model
fitting over a wide range of parameter space have been as functions of
the IR intensity at 24, 70, and 160$\mu$m.

We only need a linear tracer of dust mass to solve for $\alpha_{\rm
  CO}$, the normalization affects $\delta_{\rm GDR}$ but not
$\alpha_{\rm CO}$. While values of $\delta_{\rm GDR}$ that we find are
interesting on their own, the overall normalization of the
\citet{DRAINE07A} models do not affect our results for $\alpha_{\rm
  CO}$. Before settling on the \citet{DRAINE07A} models, we also used
modified blackbody fits with a range of emissivities to estimate the
dust opacity, $\tau_{160}$. We obtained very similar results for
$\alpha_{\rm CO}$ using that approach, in the end preferring the
\citet{DRAINE07A} models mainly for their more straightforward
handling of the 70$\mu$m band, which can include significant
out-of-equilibrium emission. When we derive the uncertainties
  associated with our assumptions (Section \ref{sec:solution} and
  Appendix \ref{sec:uncertainty}) we adopt either a modified blackbody
  or the \citet{DRAINE07A} models with equal probability. When using a
  modified blackbody fit, we include a variable fraction of 70$\mu$m
  emission from an out of equilibrium population and emissivity power
  law index in the calculation.

{\em M31:} M~31 is quiescent compared to our other targets. It has a
very low $I_{70}/I_{160}$, implying very low radiation fields or
colder dust temperatures \citep[][]{GORDON06}.  \citet{MONALTO09}
showed that with only {\em Spitzer} data the radiation field
illuminating the dust is unconstrained. In addition to poor
constraints on the model, these low $I_{70}/I_{160}$ make ratio maps
extremely sensitive to artifacts or contamination by point sources.
M~31, of all our targets, will benefit most from the additional SED
coverage offered by {\em Herschel}. Our approach in the meantime is to
treat all points in M~31 with as though they had the median IR color
across the whole galaxy. We implement the recommendation by
\citet{DRAINE07B} that in the absence of sub-millimeter data, the
radiation field be limited to $0.7$ times the local value. This
treatment effectively assumes that value everywhere in M~31 and uses
the $I_{160}\mu$m to estimate the dust mass. We note this uncertainty
in Table \ref{tab:results}.

{\em NGC6822:} NGC 6822 shows the faintest IR emission of any of our
targets. As a result, many of the variations in $I_{70}/I_{160}$
appear driven by noise and artifacts rather than $T_{\rm dust}$
variations. \citet{GRATIER10} noted the difficulty of deriving dust
temperatures for each line of sight. Based on plots of $I_{70}$
vs. $I_{160}$, we identify two main "colors" in our data. We assign
each line of sight the median color for its group, with the cut at
$I_{70}/I_{160} = 0.6$. We note this uncertainty in Table
\ref{tab:results}.

\subsection{Diffuse \hi }
\label{sec:diffuse_hi}

Gas with different $\delta_{\rm GDR}$ may be superposed along the line
of sight, in which case our assumption of a single $\delta_{\rm GDR}$
no longer applies. In some cases, we expect there to be a diffuse
\hi\ component with little or no associated dust along the line of
sight. Such a component requires that we introduce an additional term
into Equation \ref{eq:model} so that $\Sigma_{\rm gas}$ remains finite
once $\Sigma_{\rm dust} \sim 0$.

The simplest example of this is an edge-on spiral galaxy with a strong
radial gradient in $\delta_{\rm GDR}$. M~31 is inclined and does show
a gradient in $\delta_{\rm GDR}$ \citep{NIETEN06}. The pile-up of many
different radii along a single line of sight will be most severe along
the minor axis and Equation \ref{eq:model} does not appear to describe
these data as well as those along the major axis. Because we have
plenty of data, we simply exclude $60\arcdeg$ around the minor axes
from our analysis. This improves the stability of our solution. No
other correction appears necessary.

The SMC has an extended distribution of \hi\ that may also be very
elongated along the line of sight. In this \hi\ envelope and away from
the main star-forming regions (in the "Wing" and "Tail"), $\delta_{\rm
  GDR}$ is observed to be high \citep{BOT04,LEROY07,GORDON09}. Similar
results are found for other dwarf irregulars
\citep{WALTER07,DRAINE07B}. Given the very high column densities found
in the SMC it is plausible that low $\delta_{\rm GDR}$ \hi\ lies along
the line of sight.

To account for an envelope of dust-poor \hi , we subtract a diffuse
component from the SMC \hi\ map before fitting for $\alpha_{\rm
  CO}$. We estimate this component via an OLS bisector fit relating
$\Sigma_{\rm dust}$ and $\Sigma_{\rm HI}$ at low $\Sigma_{\rm dust}$,
where \hi\ is likely to represent most of the gas. This line implies a
value of $\Sigma_{\rm HI}$ where $\Sigma_{\rm dust} = 0$, which we
take as our estimate of the diffuse \hi . We estimate a diffuse
component of $\Sigma_{\rm HI} = 40$, $20$, and
$60$~M$_{\odot}$~pc$^{-2}$ for the western, northern, and eastern
parts of the SMC. Similar fits to the other targets suggest a diffuse
\hi\ component with magnitude below $10$~M$_\odot$~pc$^{-2}$, usually
consistent with zero. Our uncertainty estimates include a 20\%
($1\sigma$) uncertainty on this diffuse component in the SMC and $\pm
5$~M$_\odot$~pc$^{-2}$ in the case of galaxies other than the SMC.

This subtraction of diffuse \hi\ differs from \citet[][though the
  latter applied this approach in some of the
  analysis]{LEROY07,LEROY09}. The effect on $\alpha_{\rm CO}$ is
largest in the "east" section of the SMC, which is embedded in the
mostly diffuse Wing.

\subsection{Solution}
\label{sec:solution}

The Appendix lays out our method of solution and uncertainty estimates
in detail. We identify the best-fit $\alpha_{\rm CO}$ for each data
set as the value that minimizes the point-to-point (RMS) scatter in
$\log_{10} \delta_{\rm GDR}$. We estimate the uncertainty from three
sources: 1) statistical noise, 2) robustness to removal of individual
data, and 3) assumptions. These are reported in Table
\ref{tab:results} and we take the overall uncertainty to be the sum of
all three terms in quadrature.

\subsection{Limitations of the Model}

Our model has several important limitations. It cannot recover a
pervasive, CO-free H$_2$ component like that suggested for the LMC by
\citet{BERNARD08}. To derive such a component we would need to make
strong assumptions about $\delta_{\rm GDR}$. We derive $\alpha_{\rm
  CO}$ for H$_2$ associated with CO emission on $\sim 100$~pc scales,
essentially $\alpha_{\rm CO}$ for GMCs and their envelopes. Based on
UV spectroscopy \citep[e.g.,][]{TUMLINSON02} we consider a pervasive
H$_2$ phase unlikely, but our experiment makes no test of this idea
one way or the other.

We assume that $\delta_{\rm GDR}$ does not vary between the atomic and
molecular ISM. In fact, observations and theory suggest the
$\delta_{\rm GDR}$ does correlate with density. The depletion of heavy
elements from the gas phase increases with increasing density
\citep{JENKINS09}, though absorption measurements cannot probe to very
high densities. Meanwhile, accounting for the observed dust abundance
appears to require buildup of dust in molecular clouds
\citep{DWEK98,ZHUKOVSKA08,DRAINE09}. A lower $\delta_{\rm GDR}$ in
dense, molecular gas directly, linearly scales our derived
$\alpha_{\rm CO}$. Dust already accounts for $\sim 50\%$ of the
relevant heavy elements, so this effect cannot exceed a factor of
$\sim 2$ in magnitude. In fact, the effect should be even less severe
because we already study mainly the peaks of the gas distribution,
where we expect \hi\ to also have high densities.

Our dust modeling also implicitly assumes that the dust emissivity
does not vary between the atomic and molecular gas. Observations of
suggest that dust properties do change between the diffuse and dense
ISM, with an enhanced emissivity in dense gas. The best evidence for
this comes from a $\sim 30$--$50\%$ increase in far-IR optical depth
relative to optical extinction at high columns \citep[$\sim 1$~mag,
  e.g.,][]{ARCE99,DUTRA03,CAMBRESY05}. Direct comparison of virial
masses to submillimeter emission suggests a similar enhancement
\citep{BOT07}. The origin of this enhanced emissivity is thought to be
a change in the size distribution of dust, with ``fluffy,'' low-albedo
grains created in dense environments
\citep{DWEK97,STEPNIK03,CAMBRESY05}. As with $\delta_{\rm GDR}$
variations, emissivity variations directly scale the derived
$\alpha_{\rm CO}$.

Our best estimate is that emissivity and $\delta_{\rm GDR}$ variations
bias our measured $\alpha_{\rm CO}$ high by a factor of $\sim
1.5$--$2.0$. Because this estimate distills a variety literature
results, none of them definitive, applying such a correction after the
fact would simply confuse our results. More importantly, we have no
handle on how these factors vary with environment, though we expect
weaker effects at low metallicity, where shielding is weaker. We
highlight a quantitative understanding of $\delta_{\rm GDR}$ and dust
emissivity vary with environment as the most important systematics
that must be addressed to improve dust-based derivations of
$\alpha_{\rm CO}$.

Finally, our data cannot rule out a pervasive population of very cold
dust. We consider this unlikely and neglect it throughout the paper
\citep[e.g., see][]{DRAINE07B}. If such a population exists, the {\em
  Herschel} Space Telescope will identify it and our picture of dust
will change dramatically over the next few years. However, we note
that preliminary results suggest that dust masses are not dramatically
affected by the inclusion of {\em Herschel} data \citep[][]{GORDON10}.
Other systematics in using dust to trace \htwo\ have been discussed at
length elsewhere \citep[][with references to many other works
  therein]{ISRAEL97B,ISRAEL97,SCHNEE05,LEROY07} and we do not repeat
them here. Our best estimate is that none of these represent a major
concern. Almost all will be addressed by the inclusion of
long-wavelength {\em Herschel} data (and are already improved relative
to IRAS by using {\em Spitzer}'s 160$\mu$m band).

\section{Results}
\label{sec:results}

\begin{deluxetable*}{l c c c c c c c l}
%\begin{deluxetable}{l c c c c c c c l}
  %\rotate
  \tabletypesize{\small} 
  \tablewidth{0pt} 
  \tablecolumns{9}
  \tablecaption{\label{tab:results} Dust-Based $\alpha_{\rm CO}$ for Local
  Group Galaxies}
  \tablehead{\colhead{System} &
    \colhead{$\alpha_{\rm CO}$\tablenotemark{a}} & 
    \multicolumn{4}{c}{Uncertainty\tablenotemark{b}} & 
    \colhead{Correlation\tablenotemark{c}} &
    \colhead{Scatter\tablenotemark{d}} &   
    \colhead{Notes} \\ 
    & & Stat. & Rob. & Assump. & Tot. & & ($\log_{10} \delta_{\rm GDR}$) & 
  }
  \startdata
  M~31 & & \\
  ... inner & 2.1 & $0.14$ & $0.03$ & $0.26$ & $0.29$ & $0.68 \pm 0.02$ & 0.21 & $U$ unconstrained \\
  ... south & 10.0 & $0.05$ & $0.01$ & $0.12$ & $0.13$ & $0.85 \pm 0.02$ & 0.11 & $U$ unconstrained \\
  ... north & 4.2 & $0.05$ & $0.02$ & $0.19$ & $0.20$ & $0.83 \pm 0.02$ & 0.15 & $U$ unconstrained \\
  M~33 & & \\
  ...inner & 8.4 & $0.07$ & $0.03$ & $0.15$ & $0.17$ & $0.76 \pm 0.03$ & 0.08 & \\
  ...outer & 5.0 & $0.09$ & $0.05$ & $0.17$ & $0.20$ & $0.76 \pm 0.03$ & 0.10 & \\
  LMC & 6.6 & $0.04$ & $0.02$ & $0.14$ & $0.14$ & $0.84 \pm 0.01$ & 0.14 & \\
  NGC 6822 & 24 & $0.13$ & $0.13$ & $0.26$ & $0.31$ & $0.56 \pm 0.07 $ & 0.18 & averaged colors \\
  SMC & & \\
  ... west & 67 & $0.06$ & $0.02$ & $0.08$ & $0.10$ & $0.91 \pm 0.04$ & 0.09 & subtracted $\Sigma_{\rm HI}=40$~M$_{\odot}$~pc$^{-2}$~\tablenotemark{e} \\
  ... east & 53 & $0.08$ & $0.04$ & $0.44$ & $0.45$ & $0.95 \pm 0.11$ & 0.07 & subtracted $\Sigma_{\rm HI}=60$~M$_{\odot}$~pc$^{-2}$~\tablenotemark{e} \\
  ... north & 85 & $0.16$ & $0.05$ & $0.11$ & $0.20$ & $0.47 \pm 0.06$ & 0.15 & subtracted $\Sigma_{\rm HI}=20$~M$_{\odot}$~pc$^{-2}$~\tablenotemark{e} \\
  \enddata
  \tablenotetext{a}{Best fit. Units of M$_\odot$~pc$^{-2}$ (K km s$^{-1}$)$^{-1}$
    including helium.}
\tablenotetext{b}{Estimated $1\sigma$ uncertainty in $\log_{10} \alpha_{\rm
    CO}$ for each source of uncertainty. ``Stat.'' --- statistical
  (from Monte Carlo),
  ``Rob.'' --- robustness to removal of data (from bootstrapping),
  ``Assump.'' --- variation of assumptions.}
  \tablenotetext{c}{Linear correlation coefficient relating
    $\Sigma_{\rm gas}$ and $\Sigma_{\rm dust}$ for best-fit
    $\alpha_{\rm CO}$. Uncertainty gives $1\sigma$ scatter for
    random re-pairings of data.}
  \tablenotetext{d}{$1\sigma$ scatter in $\log_{10} \delta_{\rm GDR}$
    in dex for best-fit $\alpha_{\rm CO}$.}
  \tablenotetext{e}{Based on fit of $\Sigma_{\rm HI}$ to $\Sigma_{\rm
      dust}$ at low $\Sigma_{\rm dust}$.}
%\end{deluxetable}
\end{deluxetable*}

\begin{figure*}
\plotone{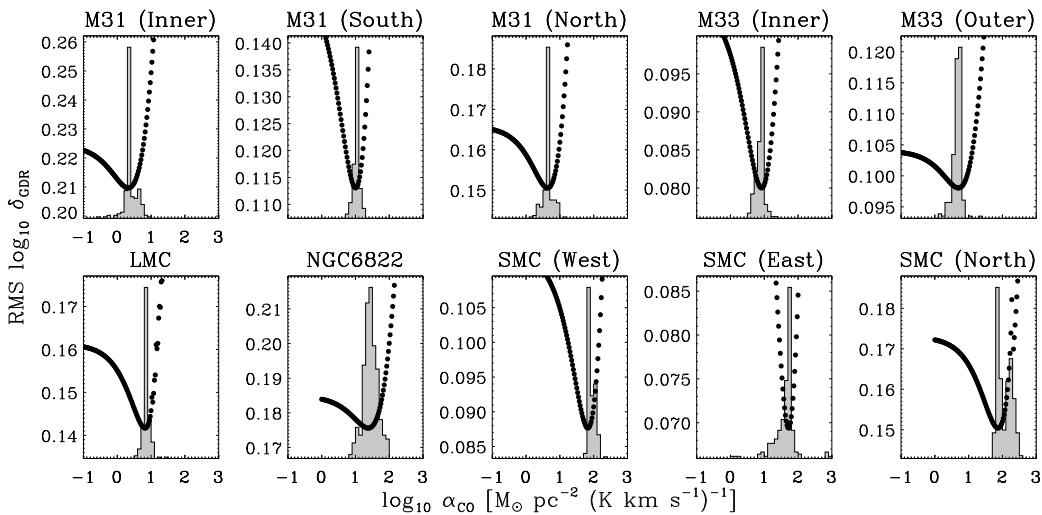}
\caption{\label{fig:chisq_vs_alpha} Scatter in $\log_{10} \delta_{\rm
    GDR}$ as a function of $\alpha_{\rm CO}$; each dot shows a
  calculation for a trial $\alpha_{\rm CO}$. The shaded histograms in
  the background show the distribution of best-fit $\alpha_{\rm CO}$
  achieved across bootstrapping and variation of assumptions; they
  give some indication of the likelihood of a given $\alpha_{\rm CO}$
  describing the data.}
\end{figure*}

\begin{figure*}
\plotone{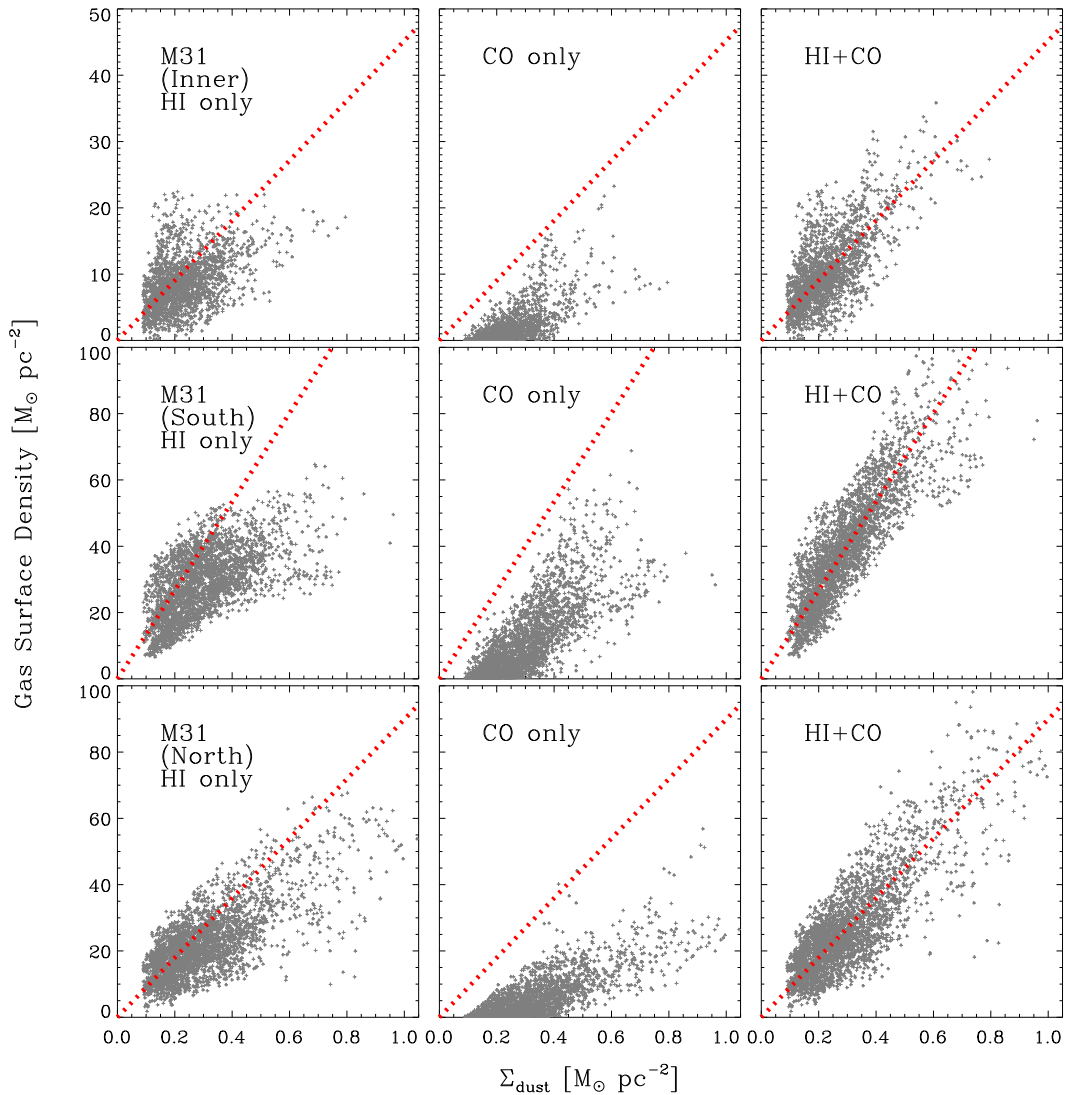}
\caption{\label{fig:scatter_plots_1} Scatter plots showing
  $\Sigma_{\rm gas}$ as a function of $\Sigma_{\rm dust}$ for
  M~31. The columns show the relationship between dust ($x$-axis) and
  ({\em left}) \hi , ({\em middle}) H$_2$ derived from CO using our
  best-fit $\alpha_{\rm CO}$, and ({\em right}) total (\hi\ + H$_2$)
  gas. Our best-fit $\alpha_{\rm CO}$ yields reasonable linear
  scalings between total gas and dust in each system (the red line
  shows the median $\delta_{\rm GDR}$), usually a clear improvement on
  the relationship between dust and either phase alone.}
\end{figure*}

\begin{figure*}
\plotone{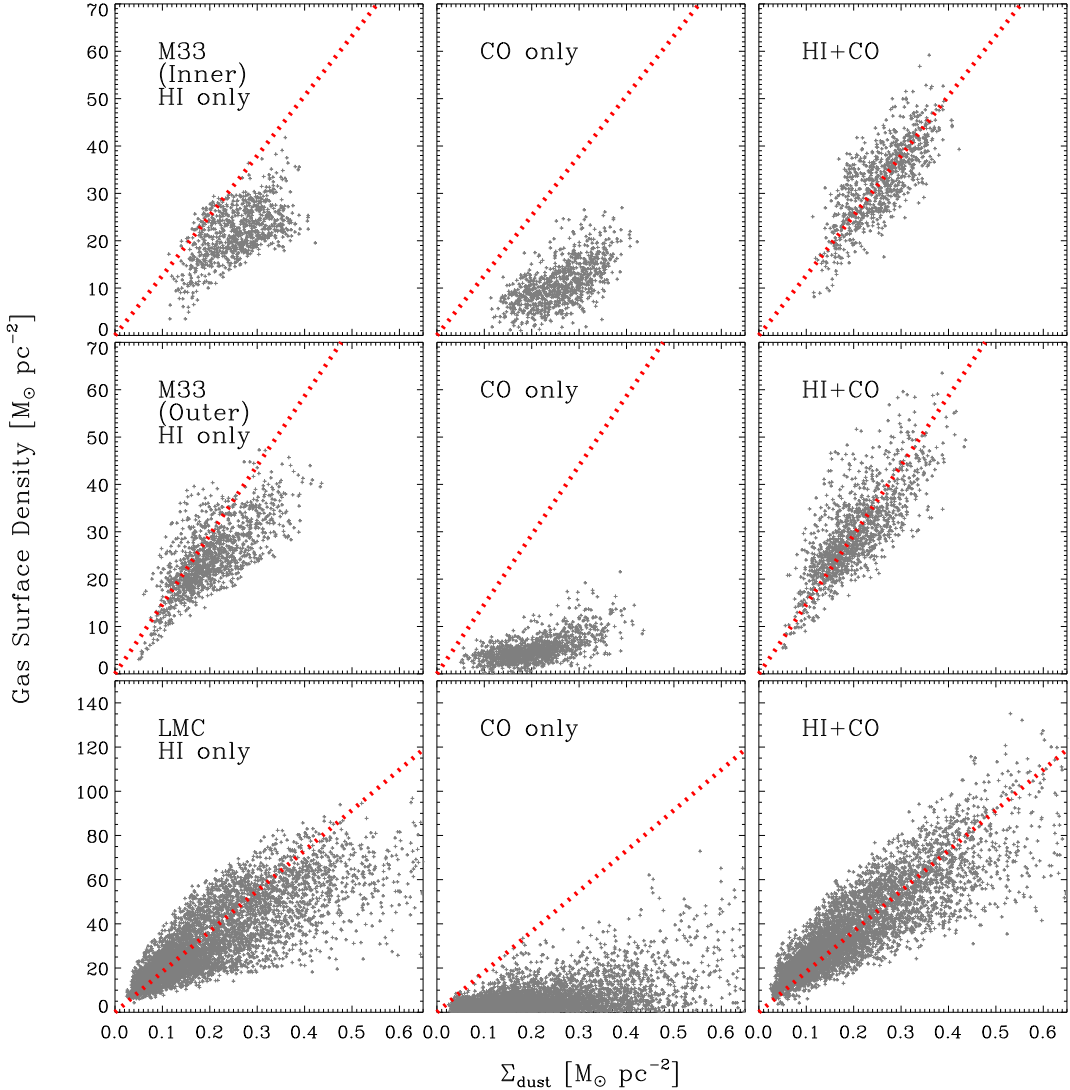}
\caption{\label{fig:scatter_plots_2} As Figure
  \ref{fig:scatter_plots_1} for M~33 and the LMC.}
\end{figure*}

\begin{figure*}
\plotone{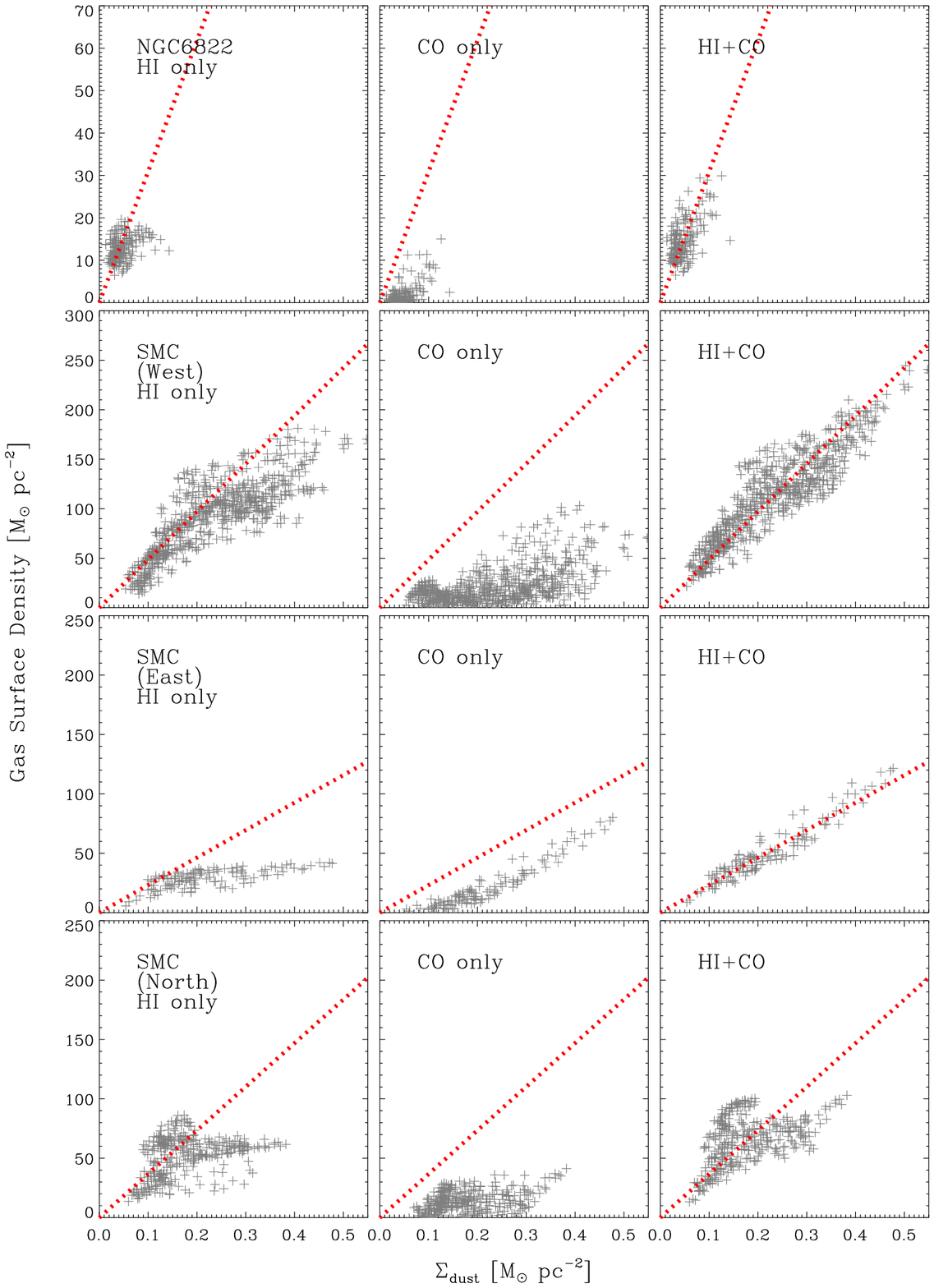}
\caption{\label{fig:scatter_plots_3} As Figure
  \ref{fig:scatter_plots_1} for NGC~6822 and the SMC.}
\end{figure*}

Figure \ref{fig:chisq_vs_alpha} plots scatter in $\log_{10}
\delta_{\rm GDR}$ as a function of $\alpha_{\rm CO}$ for our
targets. In the background, a normalized histogram indicates the
distribution of best-fit $\alpha_{\rm CO}$ derived across the
uncertainty exercises \#2 (bootstrapping) and \#3 (variation of
assumptions). We find clear minima in each data set, which we report
in Table \ref{tab:results} with associated uncertainties. We also
quote the linear correlation coefficient between gas and dust and the
scatter in $\log_{\rm 10} \delta_{\rm GDR}$ for the best-fit
$\alpha_{\rm CO}$. Both quantities indicate the degree to which our
model describes the system.

The scatter plots in Figures \ref{fig:scatter_plots_1},
\ref{fig:scatter_plots_2}, and \ref{fig:scatter_plots_3} show gas
($y$-axis) as a function of dust ($x$-axis) for each target. In the
left column $\Sigma_{\rm gas}$ is calculated from \hi\ alone, the
middle column shows $\Sigma_{\rm gas} = \alpha_{\rm CO} I_{\rm CO}$
for our best-fit $\alpha_{\rm CO}$, and the right column shows total
gas ($\Sigma_{\rm gas} = \Sigma_{\rm HI} + \alpha_{\rm CO} I_{\rm
  CO}$). In each case, the last column is a better match to a line
through the origin (i.e., a single $\delta_{\rm GDR}$) than the first
two.

The southern part of M~31 exemplifies the signal that we look
for. \hi\ correlates well with dust at low $\Sigma_{\rm dust}$. At the
high end, $\Sigma_{\rm dust}$ increases without a corresponding
increase in \hi\ (left panel). CO correlates well with $\Sigma_{\rm
  dust}$ at high $\Sigma_{\rm dust}$ but drops to low values (or zero)
while there is still significant dust, so that the $x$-intercept of
the CO-$\Sigma_{\rm dust}$ relationship is not zero (middle
panel). The gas becomes mostly molecular above a certain $\Sigma_{\rm
  dust}$ and is mostly atomic below this. Our best $\alpha_{\rm CO}$
stitches these regimes together, allowing a single $\delta_{\rm GDR}$
to span all data.

\subsection{Comments on Solutions}
\label{sec:comments}

Before arriving at these results we varied our methodology
significantly. We tried different dust treatments, methods of
solution, sampling regions, and treatments of diffuse \hi . Most
variations yield similar results to what we present here but a few
cases are worth comment.

First, the northern part of the SMC is not well-described by our model
even after we excluded the high CO-to-IR clouds. Several distinct
groups of data appear in the scatter plots for this region, suggesting
local variations in $\delta_{\rm GDR}$ or $\alpha_{\rm CO}$. The data
for this region are already good \citep[][]{MUELLER10}, so improved
modeling --- multiple dust and {\sc Hi} components, local variations
in $\alpha_{\rm CO}$ and $\delta_{\rm GDR}$ --- seems like the most
acute need. In the meantime our solution minimizes $\delta_{\rm GDR}$
and yields a result consistent with the other SMC regions.

NGC~6822 yields reasonable solutions but is somewhat unstable to
choice of methodology.  Adjusting the dust treatment, background
subtraction, or weighting can change the best-fit $\alpha_{\rm CO}$ by
$\sim 50\%$. This instability results mostly from the low signal to
noise ratio and dynamic range in the data. These, in turn, are low
because of the large distance NGC~6822 compared to the SMC or the LMC,
which makes observations challenging. The resulting low intensities
also make confusion with the Milky Way an issue
\citep{CANNON06}. Improved resolution from {\em Herschel} or ALMA
should allow a greater range of $\Sigma_{\rm dust}$ as individual
clouds are resolved. This will enable a stronger constraints on
$\alpha_{\rm CO}$.

The inner part of M~31 is also somewhat unstable to choice of
methodology --- especially region definition and fitting
method. Best-fit solutions can range from $\alpha_{\rm CO} \sim
1$--$5$. The most likely cause is variations in $\delta_{\rm GDR}$ and
$\alpha_{\rm CO}$ within the region studied --- the data do not appear
to be a plane in CO-\hi -dust space. Both improved modeling and better
IR SED coverage will help this case.

We emphasize that our approach to the SMC is conservative. We subtract
a diffuse \hi\ foreground, correct for high optical depth \hi , and
solve for $\delta_{\rm GDR}$ in the star-forming regions only ---
rendering us insensitive to any pervasive CO-free
\htwo\ component. All of these choices lower $\alpha_{\rm CO}$
bringing it closer to the other galaxies, but the SMC still exhibits
notably high $\alpha_{\rm CO}$. The more straightforward approaches
employed in \citet{ISRAEL97B, LEROY07,LEROY09} yield even higher
$\alpha_{\rm CO}$.

\subsection{$\alpha_{\rm CO}$ vs. Metallicity}

\begin{figure*}
\plottwo{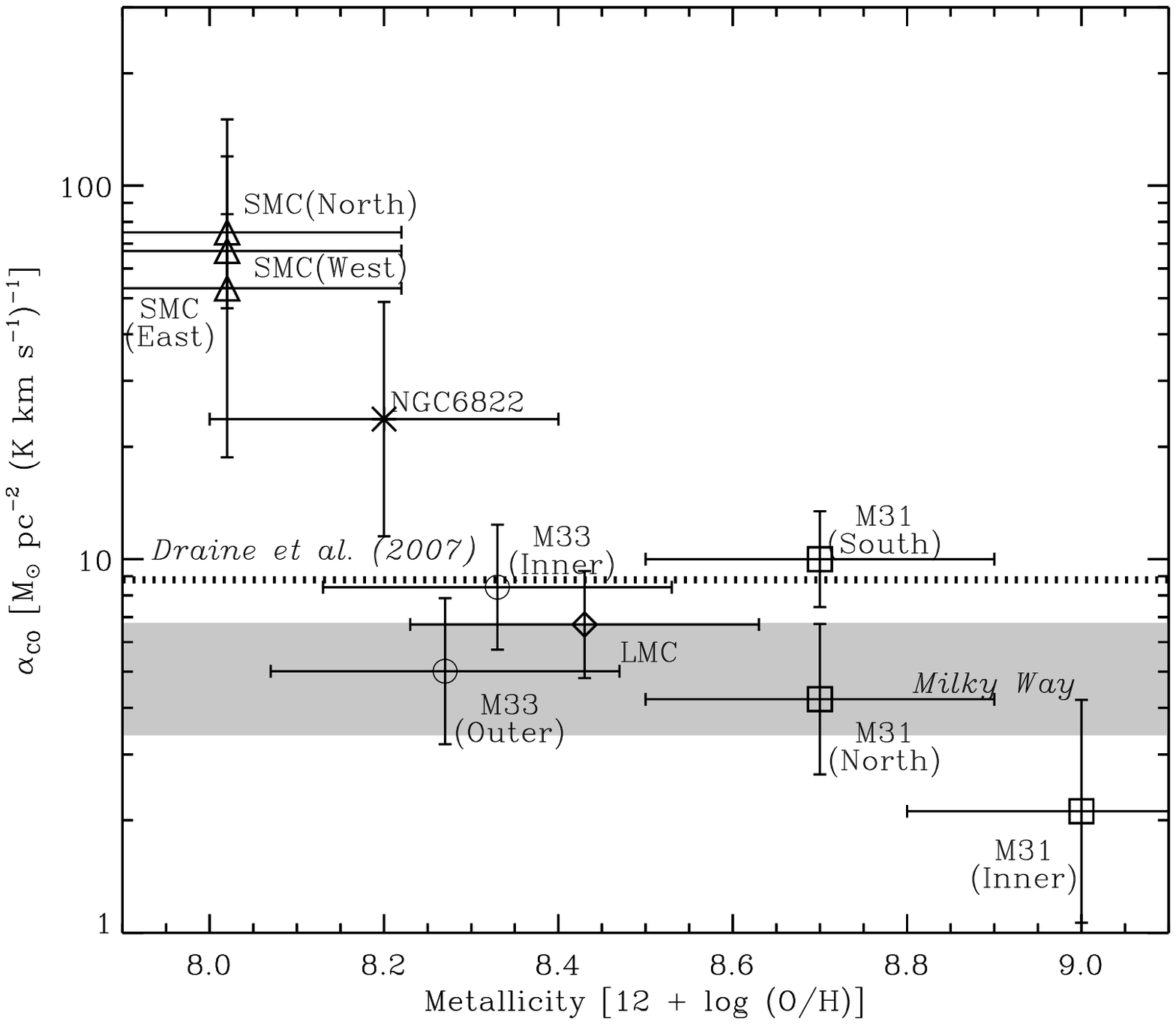}{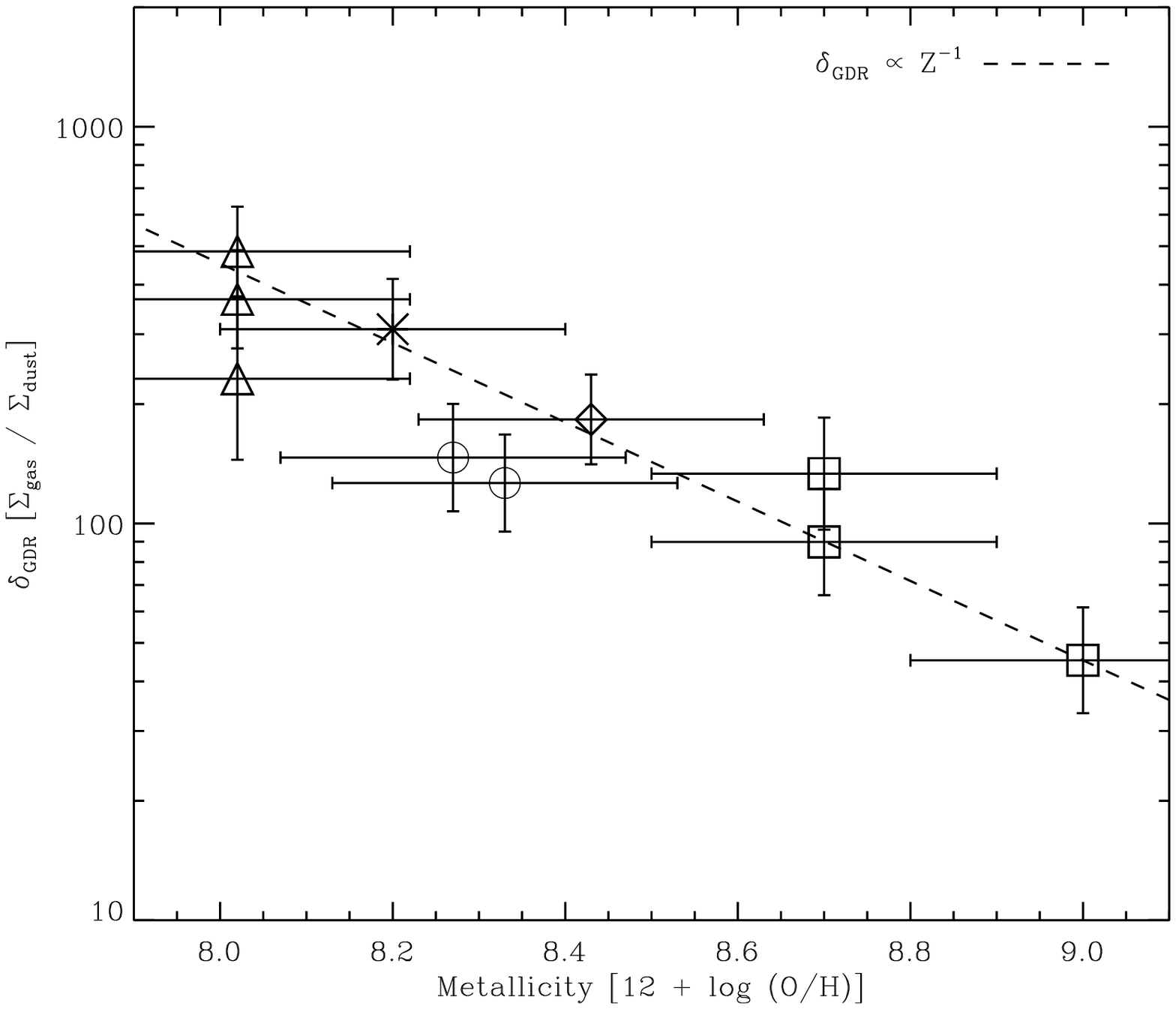}
\caption{\label{fig:alpha_vs_metals} {\em (Left)} $\alpha_{\rm CO}$ as
  a function of metallicity.  The gray region shows the range of
  commonly-used $\alpha_{\rm CO}$ for the Milky Way and the dashed
  line indicates the value argued for by \citet{DRAINE07B} studying
  integrated photometry of SINGS galaxies. {\em (Right)} The
  gas-to-dust ratio ratio $\delta_{\rm GDR}$ as a function of the same
  metallicities. The dashed line indicates a linear scaling.}
\end{figure*}

Figure \ref{fig:alpha_vs_metals} shows $\alpha_{\rm CO}$ (left) and
$\delta_{\rm GDR}$ (right) as a function of metallicity. Table
\ref{tab:results} gives the adopted metallicity for each target, which
unfortunately represent a major source of uncertainty. Literature
values span several tenths of a dex for most of our systems and
internal variations in 12+$\log ({\rm O/H})$ often exist even where
strong gradients are absent. We show a $0.2$~dex uncertainty in
12+$\log ({\rm O/H})$ in our plots to reflect this uncertainty.

Figure \ref{fig:alpha_vs_metals} shows that $\delta_{\rm GDR}$ is a
clear function of metallicity, increasing with decreasing
metallicity. This clear, one-to-one dependence of $\delta_{\rm GDR}$
on metallicity provides a consistency check on our solutions for
$\alpha_{\rm CO}$. The data are roughly consistent with a linear
relation (i.e., a fixed fraction of metals in dust), though a fit
yields a slightly sublinear slope, $\log_{10} \delta_{\rm GDR} =
\left( 9.4 \pm 1.1 \right) - \left(0.85 \pm 0.13\right) \left[12 +
  \log ({\rm O/H})\right]$. Our focus on star-forming regions likely
biases us towards high $\delta_{\rm GDR}$; studies of all material
find a steeper relation \citep{LISENFELD98,MUNOZMATEOS09}.

In the left panel of Figure \ref{fig:alpha_vs_metals} we plot the
best-fit $\alpha_{\rm CO}$ as a function metallicity. A gray area
shows the commonly accepted range of values for the Milky Way
\citep{BLOEMEN86,SOLOMON87,STRONG96,DAME01,HEYER09}. A dashed line
shows the $\alpha_{\rm CO}$ argued by \citet{DRAINE07B} to offer the
best consistency between dust and gas measurements in SINGS
galaxies. The molecular ring in M~31, both parts of M~33, and the LMC
span a factor of $\sim 2$--$3$ in metallicity but show little clear
gradient in $\alpha_{\rm CO}$. All three appear broadly consistent
with the Milky Way. We find higher $\alpha_{\rm CO}$ in NGC~6822 and
the SMC, but the rise relative to the higher metallicity systems much
steeper than linear. Figure \ref{fig:alpha_vs_metals} therefore
suggest a jump from "normal" to "high" $\alpha_{\rm CO}$ at
metallicity $\sim 1/4$ the solar value rather than a steady dependence
of $\alpha_{\rm CO}$ on metallicity. At the high metallicity end, the
inner part of M~31 does suggest a low $\alpha_{\rm CO}$. It is unclear
how much weight to ascribe to this point and whether the low
$\alpha_{\rm CO}$ is primarily a product of metallicity: the solution
is somewhat unstable and a number of other conditions differ between
the bulge of M~31 and our other targets. However, the point does
highlight that lower-than-Galactic $\alpha_{\rm CO}$ have often been
observed, especially in extreme, dust-rich systems
\citep[e.g.,][]{DOWNES98}.

\section{Discussion}
\label{sec:discussion}

First, we emphasize a very basic result: combining \hi , CO, and IR
emission we obtain reasonable solutions for $\alpha_{\rm CO}$ across
the Local Group. In M~31, M~33, and the LMC $\alpha_{\rm CO}$ appears
consistent, within the uncertainties, with the Milky Way. In most
targets our $\alpha_{\rm CO}$ agree well with previous determinations
using other methods. Despite its simplicity and the potential
importance of systematic effects, the approach outlined here produces
consistent results. A lack of quantitative constraints on how the
  emissivity and $\delta_{\rm GDR}$ change between the atomic and
  molecular ISM (ideally measured at several metallicities) remains
  the most significant obstacle to derive robust absolute values of
  $\alpha_{\rm CO}$ from this approach. We cite circumstantial
  evidence that these effects combined may cause us to overestimate
  $\alpha_{\rm CO}$ everywhere by a factor of $\sim 1.5$--$2$ but
  emphasize the need for better constraints to improve the precision
  of this approach. In the meantime we have, we believe, an internally
  robust measurement of $\alpha_{\rm CO}$ spanning an order of
  magnitude in metallicity.

\subsection{CO-dark Gas At Low Metallicities}

Does a rapid increase in $\alpha_{\rm CO}$ over a narrow range in
metallicity make physical sense? The metallicity dependence of
$\alpha_{\rm CO}$ is likely driven by "CO-dark" \htwo\ at low
extinctions. In regions with low metallicity a significant mass of
\htwo\ is found in the outer parts of clouds where the carbon is
mostly C$^{\rm +}$ rather than CO
\citep[e.g.,][]{MALONEY88,ISRAEL97,BOLATTO99}. Both \citet{GLOVER10}
and \citet{WOLFIRE10} recently studied this problem by modeling
molecular clouds. They find that the dust extinction is the key
parameter determining the fraction of CO-dark gas or $\alpha_{\rm
  CO}$. For clouds with mean extinction ${A_{\rm V}} \sim 8$~mag, such
as those in the Milky Way, \citet{WOLFIRE10} find a fraction of
CO-dark gas $f_{DG}\sim 30\%$, in rough agreement with several
observations of nearby clouds \citep{GRENIER05,ABDOXCO}. At
intermediate metallicities the CO-dark \htwo\ is not dominant, so that
even a significant fractional change does not impact $\alpha_{\rm CO}$
very much. For example, in the calculations by \citet{WOLFIRE10} going
from the metallicity of M~31 to that of the LMC causes a doubling of
the CO-dark gas fraction, from $f_{DG}\approx20\%$ to
$f_{DG}\approx40\%$, but the corresponding effect on $\alpha_{\rm CO}$
is only a factor of $\sim 1.3$, easily within the uncertainties of our
study.  As the mean extinction through the cloud decreases due to the
effects of metallicity on the gas-to-dust ratio the increase in the
fraction of CO-dark \htwo\ makes it the dominant molecular component,
at which point $\alpha_{\rm CO}$ is very rapidly driven upward.

Our measurements suggest that CO-dark gas becomes an important
component in the metallicity range $12+\log ({\rm O/H}) \sim
8.2-8.4$. With such a small sample we cannot be sure that this is a
general result, but it is in good agreement with the expectation from
\citet{WOLFIRE10}. Above these metallicities variations in the
fraction of CO-dark gas will still exist, but their influence on the
value of $\alpha_{\rm CO}$ will be small.  Excitation of the molecular
gas (particularly due to temperature variations) will be likely the
dominant factor setting $\alpha_{\rm CO}$ in molecule-rich
systems. This is probably the cause of the low $\alpha_{\rm CO}$
observed in LIRGs and ULIRGs \citep[e.g.,][]{DOWNES98}, which have
solar or slightly subsolar metallicities
\citep[e.g.,][]{RUPKE08}. Within galaxies the radiation field incident
on the molecular gas may also play a role, a fact emphasized by
\citet{ISRAEL97B}. Unfortunately, the size scales involved make this
extragalactic measurement difficult. While \citet{ISRAEL97B} found
strong quantitative support for the influence of the radiation field
on $\alpha_{\rm CO}$, \citet{PINEDA09} and \citet{HUGHES10} recently
failed to detect this effect in carefully controlled experiments in
the LMC.

\subsection{Comparison With Literature}

\begin{figure*}
\plotone{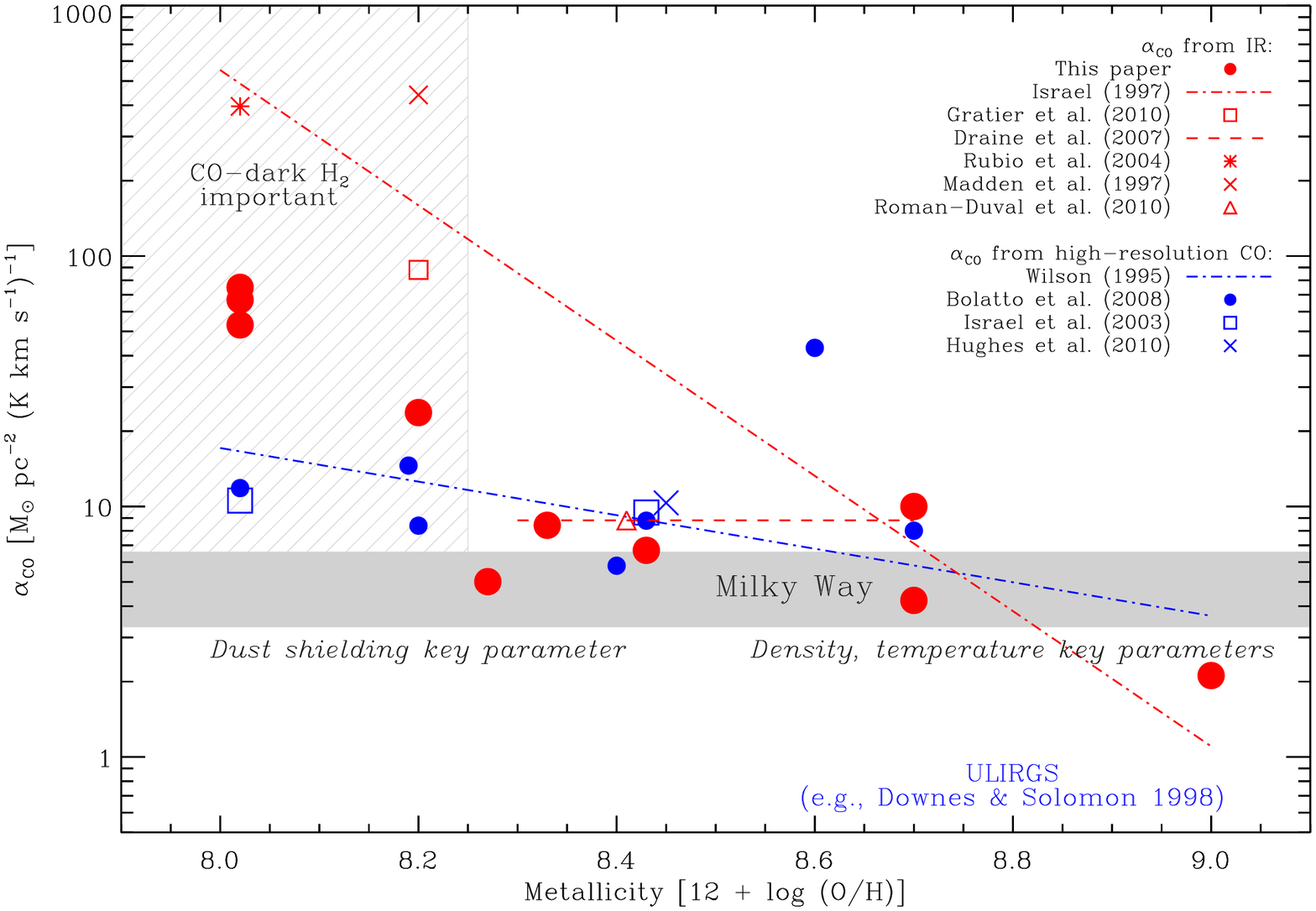}
\caption{\label{fig:fancy_plot} $\alpha_{\rm CO}$ as a function of
  metallicity. Blue measurements show $\alpha_{\rm CO}$ from virial
  mass calculations using high-resolution ($\lesssim 30$~pc FWHM) CO
  mapping. Red measurements show $\alpha_{\rm CO}$ from infrared
  observations. The ``ULIRGS'' label indicates roughly the region of
  parameter space occupied by the dense, excited gas in merger-induced
  starbursts \citep{DOWNES98}.}
\end{figure*}

Many authors have measured $\alpha_{\rm CO}$ using a variety of
techniques. We will not attempt to summarize the literature, but focus
on comparisons to two sets of measurements: 1) previous applications
of IR-based techniques; and 2) high spatial resolution measurements of
dynamical masses using CO emission.

We measure high $\alpha_{\rm CO}$ in the SMC and NGC~6822. This agrees
with a larger trend in which IR photometry and spectroscopy suggest
high $\alpha_{\rm CO}$ in the range $12+\log ({\rm O/H}) \lesssim
8.0$--$8.2$. \citet{ISRAEL97B} saw this in a number of irregulars.
\citet{RUBIO04}, \citet{LEROY07}, and \citet{LEROY09} found similar
results in the SMC, though as we note in Section \ref{sec:comments} we
actually solve for a somewhat lower $\alpha_{\rm CO}$ than these
studies --- a fact we attribute to our focus on deriving $\delta_{\rm
  GDR}$ from the \hi\ immediate associated with the star-forming
region. \citet{GRATIER10} found the same high $\alpha_{\rm CO}$ for
NGC~6822 using several approaches. Using far-infrared spectroscopy
\citet{MADDEN97} found indications of high $\alpha_{\rm CO}$ in IC~10,
a Local Group galaxy with metallicity similar to NGC~6822 (we do not
include IC~10 in this study because it lies near the Galactic plane
and has considerable IR foregrounds). \citet{PAK98} reached similar
conclusions in the SMC. Dust continuum modeling by \citet{BERNARD08}
and \citet{ROMANDUVAL10} found a mixed picture in the LMC, also
consistent with our findings.  Note that in contrast with this broad
agreement, recent work on diffuse lines of sight in the Milky Way
\citep{LISZT10} suggests that the ratio of CO brightness to
\htwo\ column density is not a strong function of column density.

Figure \ref{fig:fancy_plot} compares $\alpha_{\rm CO}$ as a function
of metallicity between this study and the literature. The points in
red indicate IR-based measurements.  In detail, our measurements
(circles) yield {\em lower} $\alpha_{\rm CO}$ than previous IR-based
studies. We suspect that this is mainly because we solve for
$\alpha_{\rm CO}$ without assuming $\delta_{\rm GDR}$ or measuring it
far away from the region of interest. One likely sense of systematic
variations in $\delta_{\rm GDR}$ is that $\delta_{\rm GDR}$ is likely
to be higher in the dense gas close to molecular complexes, which
  tend to reside mainly in the stellar disk, than in a diffuse,
extended \hi\ disk
  \citep[e.g.,][]{STANIMIROVIC99,DRAINE07B,MUNOZMATEOS09}. If
$\delta_{\rm GDR}$ is taken to be too high, Equation \ref{eq:model}
yields a corresponding overestimate of $\alpha_{\rm CO}$. In the
  SMC our attempt to remove a diffuse {\sc Hi} component along the
  line of sight also leads to lower $\alpha_{\rm CO}$ (Section
  \ref{sec:comments}), though it is less clear that our approach is
  correct in that case. Regardless of the cause, by simultaneously
  solving for $\alpha_{\rm CO}$ and $\delta_{\rm GDR}$ in the regions
  of interest in a uniform way across a heterogeneous sample we
  improve on literature studies of individual galaxies.

A long standing discrepancy exists between IR-based results and
high-resolution virial mass measurements based on CO
observations. Using virial masses, \citet{WILSON95},
\citet{ROSOLOWSKY03}, and \citet{BOLATTO08} all found weak or absent
trends in \xco\ as a function of metallicity. The blue points in
Figure \ref{fig:fancy_plot} show virial mass results from CO
observations with resolution better than 30~pc. The two approaches
agree up to about the metallicity of M~33 or the LMC, and then
strongly diverge in the SMC. This divergence is most easily understood
if the additional \htwo\ traced by IR lies in an extended envelope
outside the main CO emitting region \citep{BOLATTO08}. Such an
envelope can reconcile the virial mass and dust measurements and
naturally explains the scale-dependence of $\alpha_{\rm CO}$ observed
by \citet{RUBIO93B} in the SMC. These envelopes could perhaps still
have an effect on the velocity dispersion of the material inside it
(and consequently the measured virial mass) via surface pressure.
Structures with virial parameters $\alpha\leq 1$, however, are often
observed inside local molecular clouds, suggesting that at least in
some instances the velocity dispersion does not appreciably show the
impact of the surrounding material. An alternative view is argued by
\citet{BOT07,BOT10}, who observed discrepancies between dust-based
masses and virial masses even at fairly small scales. They suggest
that magnetic support becomes very strong at low metallicities,
perhaps because of higher ionization fractions inside clouds.

\section{Summary}

We combine CO, \hi , and IR measurements to solve for the CO-to-H$_2$
conversion factor, $\alpha_{\rm CO}$, in M~33, M~31, NGC~6822, the LMC
and the SMC. We estimate the dust mass from IR intensities and then
identify the $\alpha_{\rm CO}$ that produces the best linear relation
between total (\hi + \htwo ) gas and dust. We accomplish this finding
the $\alpha_{\rm CO}$ and $\delta_{\rm GDR}$ that minimize the scatter
about Equation \ref{eq:model}. We find that $\alpha_{\rm CO}$ is
approximately constant (within a factor of two) in M~31, M~33, and the
LMC, with a value $\alpha_{\rm CO} \approx 6$~\alphaunits. By contrast
NGC~6822 and the SMC, the lowest metallicity galaxies in the sample,
show a drastically higher $\alpha_{\rm CO}$, $\sim 30$ and $70$. The
resulting gas-to-dust ratio, $\delta_{\rm GDR}$, scales approximately
linear with metallicity.

We attribute the behavior of $\alpha_{\rm CO}$ to the transition from
the regime where most \htwo\ is bright in CO to a regime where CO is
mostly photodissociated and the bulk of the molecular reservoir is
CO-dark.  In our sample, this transition occurs around $12+\log ({\rm
  O/H}) \sim 8.4-8.2$. With only a limited number of systems, an
actual numerical prescription for $\alpha_{\rm CO}$ is beyond the
scope of the paper. These results agree qualitatively with a large
body of existing work using IR based techniques, though quantitatively
we find lower $\alpha_{\rm CO}$ than previous work
\citep[e.g.,][]{ISRAEL97B}, probably because we restrict our analysis
to CO-emitting regions.

\acknowledgements We thank the anonymous referee for a constructive
report and helpful suggestions. We gratefully acknowledge Elias
Brinks, John Cannon, and Fabian Walter for sharing their data on M~31
and NGC~6822. We also thank Julia Roman-Duval, Michele Thornley, Jack
Gallimore, Brian Kent, and Erik Mueller for helpful discussions and
comments on drafts. We acknowledge the use of NASA's Astrophysics Data
System Bibliographic Services and the NASA/IPAC Extragalactic Database
(NED). Support for AKL was provided by NASA through Hubble Fellowship
grant HST-HF-51258.01-A awarded by the Space Telescope Science
Institute, which is operated by the Association of Universities for
Research in Astronomy, Inc., for NASA, under contract NAS 5-26555.  The National Radio Astronomy Observatory is a facility of the National Science Foundation operated under cooperative agreement by Associated Universities, Inc. AB
acknowledges partial support from {\em Spitzer} JPL grant contract
1314022, and from NSF AST-0838178 and NSF-AST0955836.

\bibliography{/users/aleroy/bib/akl}

\begin{appendix}

\section{Method of Solution}

For a given $\alpha_{\rm CO}$, we can compute the total gas surface
density ($\Sigma_{\rm gas} = \Sigma_{\rm H2} + \Sigma_{\rm HI}$) for
each line of sight. With the corresponding $\Sigma_{\rm dust}$, each
point implies a value of $\delta_{\rm GDR}$. We have assumed (Equation
\ref{eq:model}) that a single $\delta_{\rm GDR}$ describes each data
set. A simple way to identify the $\alpha_{\rm CO}$ that best fits
this model is to try many $\alpha_{\rm CO}$ and select the one that
minimizes the scatter in $\log_{\rm 10}~\delta_{\rm GDR}$. First we
pick a set of $\alpha_{\rm CO}$ that bracket the true value. For each
$\alpha_{\rm CO}$, we calculate $\Sigma_{\rm gas}$ and combine it with
$\Sigma_{\rm dust}$ to measure the RMS scatter in $\log_{\rm
  10}~\delta_{\rm GDR}$. This yields clear minima in each data set,
which we identify as our best-fit $\alpha_{\rm CO}$.

We experimented with several other approaches, including plane fits
with errors in all three axes. The results agree with what we report
here except that the plane fit is less stable in the inner part of
M~31, NGC~6822, and the northern part of the SMC. The presence of
significant intrinsic scatter in the relation makes the weighting of
data subjective and so the plane fit is not significantly more
rigorous than using the scatter as a goodness-of-fit metric. We also
experimented with maximization of the linear and rank correlation
coefficients relating dust and gas. Most of these approaches yield
similar results. Our uncertainty estimates include substituting the
median-absolute-deviation for the RMS in the goodness of fit.

This method can fail in a data set without a good mix of \htwo\ and
\hi -dominated lines of sight. For example, in a data set with 100
overwhelmingly \hi\ lines of sight and 3 mostly molecular lines of
sight, the \hi -dominated data will drive the scatter in
$\log_{10}~\delta_{\rm GDR}$ leaving little sensitivity to
$\alpha_{\rm CO}$. In such a case we would like to weight the high
\htwo\ lines of sight more heavily. In practice, our definition of
sampling regions addresses this concern and we weight all points
equally.

\section{Uncertainties}
\label{sec:uncertainty}

\begin{figure}
\plotone{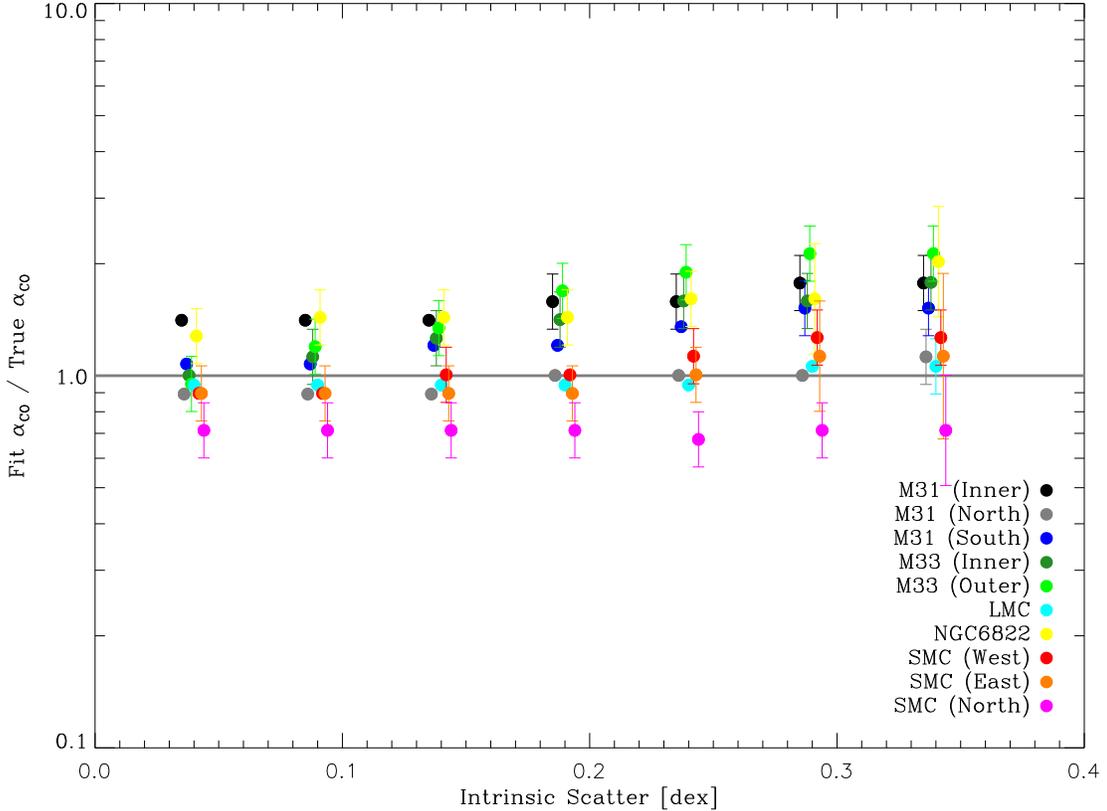}
\caption{\label{fig:bias} Fits to simulated data. The $y$-axis gives
  the mean best-fit $\alpha_{\rm CO}$ divided by the true value (known
  by construction) as a function of the intrinsic scatter in the
  dust-gas relation ($x$-axis). Error bars show the RMS scatter in the
  best-fit $\alpha_{\rm CO}$. The horizontal line at 1 shows a perfect
  match between true and best-fit $\alpha_{\rm CO}$. Different colors
  indicate different systems. For realistic intrinsic scatter
  ($\lesssim 0.2$~dex, see Table \ref{tab:results}), we recover the
  true $\alpha_{\rm CO}$ with better than $40\%$ accuracy in all
  cases.}
\end{figure}

We gauge the uncertainty in the best-fit $\alpha_{\rm CO}$ in three
ways: 1) We vary details of the calculation (the calibration and
background for each map, the dust model, and the goodness of fit
statistic) across their plausible range, repeating our solution for
each new set of assumptions. 2) We solve for $\alpha_{\rm CO}$ in a
simulated noisy data set where we know the true $\alpha_{\rm CO}$ by
construction. 3) We repeat the original solution while bootstrapping,
solving for $\alpha_{\rm CO}$ using data drawn from the original
sample (with the same number of elements) allowing repeats.  We thus
assess the sensitivity of $\alpha_{\rm CO}$ to our assumptions,
statistical uncertainty, and robustness. We bookkeep each as a
gain-style uncertainty (i.e., RMS in log $\alpha_{\rm CO}$) and then
take our overall uncertainty to be the sum of all three terms in
quadrature.

In the first test, we carry out 100 iterations. Although the
  \citet{DRAINE07A} models have been applied successfully to the SMC
  \citep{SANDSTROM10}, adopting them represents a key
  assumption. Therefore, in each iteration we randomly select with
equal probability to use either the \citet{DRAINE07A} dust models or a
modified blackbody with $\beta =1$--$2$ and assuming 50\%\ of the
70$\mu$m emission to come from out-of-equilibrium very small grain
emission, a value appropriate for the SMC
  \citep[][]{LEROY09}. We also select our goodness of fit statistic
to be with equal probability either the RMS scatter (as in the main
result) in $\log_{10} \delta_{\rm GDR}$, the fractional scatter in
$\delta_{\rm GDR}$, or the median absolute deviation in $\log_{10}
\delta_{\rm GDR}$. We randomize the the zero point of the IR maps and
the absolute calibration of each data set within their plausible
value. Finally we adjust the zero point of the \hi\ map, either
varying the assumed zero point by $\pm 20\%$ (in the SMC) or $\pm
5$~M$_\odot$~pc$^{-2}$ (in the other galaxies). In detail, we vary the
zero points of the IR maps by typically $\pm 0.1$ and $0.5$ at 70 and
160$\mu$m ($1\sigma$, this depends slightly on the target). We adjust
overall scale of the CO and \hi\ data by $\pm 15\%$ ($1\sigma$) and
the IR maps by $\pm 10\%$. When we use a modified blackbody we pick
$\beta$ randomly from the range $1$--$2$ (always using a single value)
and take a fraction $0.5 \pm 0.1$ of the 70$\mu$m emission to
  come from out of equilibrium emission and so neglect it in the
  calculation. Each realization corresponds to a reasonable set of
assumptions and yields a valid solution for $\alpha_{\rm CO}$ and
$\delta_{\rm GDR}$, but we emphasize that {\em the results presented
  in the main text represent our best estimates.}

In the second test, we take $\Sigma_{\rm HI}$ and $I_{\rm CO}$ for
each data set, assume an $\alpha_{\rm CO}$, add intrinsic scatter to
the relation, add noise, and then solve for $\alpha_{\rm CO}$. We
first take the real \hi\ and CO data and assume the best-fit values of
$\alpha_{\rm CO}$ from Table \ref{tab:results} to be correct. We
convert $I_{\rm CO}$ to $\Sigma_{\rm H2}$ and then apply the measured
$\delta_{\rm GDR}$ to derive a $\Sigma_{\rm dust}$ for each point. At
this point we have a plane in CO--\hi --dust space that is perfectly
described by a single $\alpha_{\rm CO}$ and $\delta_{\rm GDR}$. We add
lognormal scatter to each quantity (equally distributed among the
three axes) and finally apply noise to each axis. This yields a
realistic approximation of real data for which we know the true
$\alpha_{\rm CO}$. We do not know the intrinsic scatter {\em a
  priori}, so we try a range of values from 0 to 0.4 dex, applying
$1/\sqrt{3}$ times this value to each axis. We repeat the exercise
several times for each scatter, resampling the original data (allowing
repeats) and re-generating the noise.

Figure \ref{fig:bias} shows the result of this test. The $y$-axis
gives the best-fit $\alpha_{\rm CO}$ divided by $\alpha_{\rm CO}$
assumed for the simulation. The $x$-axis shows the amount of intrinsic
scatter applied when making the simulated data. Each point shows the
mean and scatter for $\sim 50$ tests. The statistical uncertainty
reported in Table \ref{tab:results} is the RMS scatter (in the log)
about the true $\alpha_{\rm CO}$ for all trials with intrinsic scatter
$< 0.2$ dex, a realistic value based on Table \ref{tab:results}. It
thus incorporates both the mild bias and scatter seen in Figure
\ref{fig:bias}.

\end{appendix}

\end{document}